\documentclass[12pt]{article}
\pdfoutput=1
\usepackage{cite}
\usepackage{jheppub}
\usepackage{amsfonts,amsmath,amssymb}
\usepackage{graphicx}
\usepackage{hyperref}

\newcommand{\be}{\begin{equation}}
\newcommand{\ee}{\end{equation}}
\newcommand{\bea}{\begin{eqnarray}}
\newcommand{\eea}{\end{eqnarray}}
\newcommand{\ba}{\begin{array}}
\newcommand{\ea}{\end{array}}
\newcommand{\ben}{\begin{enumerate}}
\newcommand{\een}{\end{enumerate}}
\newcommand{\bi}{\begin{itemize}}
\newcommand{\ei}{\end{itemize}}
\newcommand{\bc}{\begin{center}}
\newcommand{\ec}{\end{center}}
\newcommand{\bt}{\begin{table}}
\newcommand{\et}{\end{table}}
\newcommand{\btab}{\begin{tabular}}
\newcommand{\etab}{\end{tabular}}
\newcommand{\bfig}{\begin{figure}}
\newcommand{\efig}{\end{figure}}
\newcommand{\bs}{\begin{slide}}
\newcommand{\es}{\end{slide}}
\newcommand{\nn}{\nonumber}
\newcommand{\eref}[1]{(\ref{#1})}
\newcommand{\comment}[1]{}
\newcommand{\s}{\sigma}

\title{Consistency and Derangements in \\ Brane Tilings}

\author[*]{Amihay Hanany,}
\author[\dagger]{Vishnu Jejjala,}
\author[\ddagger]{Sanjaye Ramgoolam,}
\author[\mathsection]{and Rak-Kyeong Seong}

\affiliation[*]{
Theoretical Physics Group, The Blackett Laboratory, Imperial College London,\\
Prince Consort Road, London SW7 2AZ, United Kingdom
}
\affiliation[\dagger]{
Mandelstam Institute for Theoretical Physics, School of Physics, and NITheP,\\
University of the Witwatersrand, WITS 2050, Johannesburg, South Africa
}
\affiliation[\ddagger]{
School of Physics and Astronomy, Queen Mary, University of London,\\
Mile End Road, London E1 4NS, United Kingdom
}
\affiliation[\mathsection]{
School of Physics, Korea Institute for Advanced Study, Seoul 02455, Korea
}

\preprint{
\begin{flushright}
Imperial/TP/15/AH/05\\
QMUL-PH-15-24\\
KIAS-P15061
\end{flushright}
}

\emailAdd{a.hanany@imperial.ac.uk}
\emailAdd{vishnu@neo.phys.wits.ac.za}
\emailAdd{s.ramgoolam@qmul.ac.uk}
\emailAdd{rkseong@kias.re.kr}

\abstract{
Brane tilings describe Lagrangians (vector multiplets, chiral multiplets, and the superpotential) of four dimensional $\mathcal{N}=1$ supersymmetric gauge theories.
These theories, written in terms of a bipartite graph on a torus, correspond to worldvolume theories on $N$ D$3$-branes probing a toric Calabi--Yau threefold singularity.
A pair of permutations compactly encapsulates the data necessary to specify a brane tiling. We show that geometric consistency for brane tilings, which ensures that the corresponding quantum field theories are well behaved, imposes constraints on the pair of permutations, restricting certain products  
constructed from the pair to have no one-cycles. Permutations without one-cycles are known as derangements.
We illustrate this formulation of consistency with known brane tilings. Counting formulas for consistent brane tilings with an arbitrary number of chiral bifundamental fields are written down in terms of delta functions over symmetric groups.
\\
}

\begin{document}

\maketitle

\section{Introduction}
\textit{Brane tilings}~\cite{hk,fhkvw} are bipartite periodic graphs on a two-torus that describe a class of ${\cal N}=1$ supersymmetric quiver gauge theories that arise as worldvolume theories on $N$ D$3$-branes probing a toric Calabi--Yau cone over a Sasaki--Einstein base manifold ${\cal B}$.
Such field theories are dual to Type IIB string theory on AdS$_5\times {\cal B}$.
The interactions in these theories are specified by superpotentials, which have the form
\be
W(\phi_i) = W_+(\phi_i) - W_-(\phi_i) ~, \label{eq:w}
\ee
where $W_+$ and $W_-$ consist of sums of single trace multilinear gauge invariant operators.
Every chiral field $\phi_1, \ldots, \phi_d$ appears exactly once in $W_+$ and once in $W_-$.

\begin{table}[ht!!]
\centering
\begin{tabular}{c|cccccccccc}
\; & 0 & 1 & 2 & 3 & 4 & 5 & 6 & 7 & 8 & 9\\
\hline
\text{D5} & $\times$ & $\times$ & $\times$ & $\times$ & $\times$ & $\cdot$ & $\times$ & $\cdot$ & $\cdot$ & $\cdot$ \\
\text{NS5} & $\times$ & $\times$ & $\times$ & $\times$ & \multicolumn{4}{c}{--------$\Sigma$--------} & $\cdot$ & $\cdot$\\
\end{tabular}
\caption{\textit{Brane configuration for brane tilings.}
D5-branes are suspended between NS5-branes that wrap a holomorphic curve $\Sigma$.}
\label{tbranes}
\end{table}

The matter content of the gauge theory is summarized in a directed graph known as the \textit{quiver}.
It consists of nodes, each of which is associated to a $U(N)$ vector multiplet, and arrows, each of which is a bifundamental chiral multiplet.
Anomaly cancelation sets the condition that at every node in the quiver the number of incoming arrows is equal to the number of outgoing arrows.
Terms in the superpotential are gauge invariant and correspond to closed directed loops in the quiver.
By mapping clockwise oriented loops corresponding to $W_{+}$ terms to white nodes, and anti-clockwise oriented loops corresponding to $W_{-}$ terms to black nodes, the quiver and superpotential are encoded as a bipartite graph on a two-torus where edges connecting to precisely a white and a black node correspond to bifundamental fields of the quiver.
This is precisely the brane tiling which is known in the mathematics literature as a \textit{dimer}~\cite{2003math.....10326K,2003math.ph..11005K}.
(See~\cite{k,Yamazaki:2008bt} for other reviews.)

Beyond encoding the quiver and superpotential information of the four dimensional $\mathcal{N}=1$ theory, brane tilings as bipartite periodic graphs on a two-torus $T^2$ represent a brane configuration of D$5$-branes suspended between NS$5$-branes as shown in Table~\ref{tbranes}.
This brane configuration is T-dual to the D$3$-branes probing the toric Calabi--Yau threefold.
The two-torus of the brane tiling is the argument space $(\arg{(x)},\arg{(y)}) \in T^2$ of the space parameterized by $x\equiv (x^4,x^5) , y \equiv (x^6,x^7) \in (\mathbb{C}^*)^2$ in Table~\ref{tbranes}.
The NS$5$-branes wrap a holomorphic curve $\Sigma$ in $x,y$ determined by the Newton polynomial of the toric diagram of the toric Calabi--Yau threefold.
The projection of the holomorphic curve $\Sigma$ in $(\arg{(x)},\arg{(y)}) \in T^2$ gives the \textit{coamoeba}~\cite{2001math......8225M,2004math......3015M} which is limited by asymptotic boundaries in the bipartite graph that correspond to cycles along graph edges with non-trivial $T^2$ winding numbers.
These cycles are known as \textit{zig-zag paths}~\cite{hv,fhkv} and encode in the brane tiling picture the toric Calabi--Yau threefold geometry as well as the intersection between D$5$-branes and NS$5$-branes in the brane configuration.

The underlying combinatorial structure of brane tilings is revealed by the fact that we can express the $d$ bifundamental fields in the quiver and the superpotential of a brane tiling in terms of three permutation elements in the symmetric group, $\sigma_i\in S_d$, that satisfy a constraint $\sigma_B\cdot \sigma_W\cdot \sigma_\infty = 1$.
The permutation tuples $\sigma_W$ and $\sigma_B$ encode $W_+$ and $W_-$, respectively.
The third tuple, which is not independent, identifies the fields that are charged under a particular gauge group (say, all incoming).
Since bipartite graphs on a Riemann surface are \textit{dessin d'enfants},\footnote{
\textit{Dessin d'enfants} are children's drawings.
The reason that these are of interest to mathematicians is that the absolute Galois group acts faithfully on dessins~\cite{g}.}
the paper~\cite{jrr} developed this identification, introduced the permutation tuples which encode a brane tiling, and applied Belyi maps from an elliptic curve to $\mathbb{P}^1$ branched over three marked points in order to understand the structure of this class of toric superconformal field theories.\footnote{
See~\cite{acd,hr} for related work on ${\cal N}=2$ superconformal field theories;~\cite{dr} discusses the relation to Matrix models.}
In~\cite{hhjprr1}, additional techniques for constructing Belyi maps, especially in orbifold CFTs, were devised.
Brane tilings have three natural complex structure parameters associated, respectively, to the source of the Belyi map ($\tau_B$), the isoradial embedding of the brane tiling determined by $a$-maximization ($\tau_R$)~\cite{Intriligator:2003jj,Butti:2005vn}, and the special Lagrangian torus fibration on which the branes tiling lives ($\tau_G$).
These complex structure parameters do not all agree.
In~\cite{hhjprr1}, we see explicitly that $\tau_B \ne \tau_R$ in, for example, $L^{222}$, the non-chiral $\mathbb{Z}_2$ orbifold of the conifold, and in~\cite{hjr}, we note that $\tau_G \ne \tau_R$ in $L^{1b1}$ toric spaces.\footnote{The nomenclature is from~\cite{Franco:2005sm,bk,bfz}.}
The article~\cite{hhjprr2} further establishes that Seiberg dual theories have an identical $\tau_R$ up to $SL(2,\mathbb{Z})$ modular transformations, so these are nevertheless interesting features of brane tilings to study.

Given that the permutation language for describing brane tilings is extraordinarily compact and persuasive, in this note we seek to determine what further information can be extracted from knowing the tuples that encode the superpotential interactions of the fields.
In particular, not all bipartite periodic graphs on the two-torus correspond to what we call a geometrically consistent brane tiling and a well behaved\footnote{Here \textit{well behaved} means that the $R$-charges, chiral ring, and moduli space are determined by the tiling, and the corresponding Calabi--Yau singularity.} four dimensional $\mathcal{N}=1$ quiver gauge theory.
Geometrically inconsistent bipartite periodic graphs on the two-torus correspond to theories with undetermined infrared fixed points.
Applying $a$-maximization to such graphs leads, for instance, to zero $R$-charges for certain bifundamental chiral fields~\cite{hv}.
Additionally, the zig-zag paths of such geometrically inconsistent theories self-intersect and hinder us from determining the moduli space from the brane tiling.
We deduce that the geometric consistency conditions for brane tilings translate to the requirement that certain permutations constructed from $\sigma_B$ and $\sigma_W$ have no one-cycles.
These are known as \textit{derangements}.

The organization of the paper is as follows.
In Section~\ref{sec:btpt}, we review the description of brane tilings in terms of permutation tuples.
This is a brief recapitulation of the syntax developed in~\cite{jrr}.
In Section~\ref{sec:cd}, we establish the derangement conditions for geometric consistency.
Known brane tilings are consonant with this test.
Finally, in Section~\ref{sec:p}, we use the technology developed in this paper and in~\cite{jrr,hhjprr1,hhjprr2,hjr} in order to write down counting formulas for consistent brane tilings with a given number of chiral fields $d$.
We further provide a prospectus for future work that exploits the new technology.

%================================%
\section{Brane Tilings as Permutation Tuples}\label{sec:btpt}

The section recollects the structure of bipartite graphs on a two-torus.
Following~\cite{jrr}, permutation tuples are then introduced to describe brane tilings.

\subsection{Brane Tilings} \label{subec:bt}
Suppose we have a $U(N)^n$ gauge theory with $d$ matter fields and a superpotential of the form~\eref{eq:w}.
The brane tiling is a pictorial representation that consists of $n$ faces corresponding to each of the factors in the gauge group $U(N)^n$.
Edges in the brane tiling correspond to bifundamental fields.
A bifundamental field is charged under the two gauge groups which are the faces that the edge separates.
The edges connect white and black vertices in the graph.
Traversing the edges that meet the white vertices in a clockwise direction recovers the superpotential terms in $W_+$ while traversing the edges that meet at black vertices in the anti-clockwise direction recovers the superpotential terms in $W_-$.
The orientation fixes the arrow direction in the quiver and hence the assignment of fundamental and anti-fundamental representations to the fields.

We label the edges in the brane tiling with $1,\ldots,d$.
We can encode the structure of the tiling  in terms of a pair of permutations, $\sigma_B$ and $\sigma_W$, 
 in the symmetric group $S_d$.
The   permutation $\sigma_B$ contains a cycle for every black vertex, with length equal to the valency of the vertex. The numbers in the cycle correspond to the 
 the labelled edges encountered in going anti-clockwise round the vertex. Likewise, $\sigma_W$ has a cycle for every white vertex, again read by going anti-clockwise round the vertex. The cyclic property of the elements in the permutation cycle echoes the cyclic property of the trace.
The toric condition~\cite{Feng:2000mi} for the Calabi--Yau moduli space is accounted for by the fact that every edge appears precisely once as an entry in $\sigma_W$ and $\sigma_B$.

We recall from~\cite{jrr} that a brane tiling satisfies the following conditions regarding \textit{permutations and tuples}.
\bi
\item \textbf{PT-1.}
Two permutations $(\s_B, \s_W)$ and $(\s_B', \s_W')$ determine the same bipartite graph when
\be\label{ept1}
\s_B' = \gamma \s_B \gamma^{-1} ~, \qquad
\s_W' = \gamma \s_W \gamma^{-1} ~,
\ee
for some $\gamma\in S_d$.

\item \textbf{PT-2.}
The group generated by $\s_B$ and $\s_W$ is transitive.
Let us call this group $G(\s_B, \s_W)$.
The transitivity condition means that any integer $i$ from $\{1,\ldots,d\} $ can be mapped to any other
by some element of $G$.

\item \textbf{PT-3.}
Let $\s_F = (\s_B\, \s_W)^{-1} $.
Let $C_\s$ denote the number of cycles in $\s$.
By the Riemann--Hurwitz relation, the condition that we have a genus one bipartite graph is
\be
d - C_{\s_B} - C_{\s_W} - C_{\s_F} = 0 ~. \label{eq:rh}
\ee
As a consequence of anomaly cancelation and the Riemann--Hurwitz relation, the number of cycles in $\s_B$ and $\s_W$ are the same.
A bipartite graph with an equal number of black and white nodes is said to be \textit{balanced}.

\item \textbf{PT-4.}
Valencies specify the lengths of the cycles in $\s_B$ and $\s_W$.
We call these $T_{\s_B}$ and $T_{\s_W}$.\footnote{
While $C_{\s_B} = C_{\s_W}$, it need not be the case that $T_{\s_B} = T_{\s_W}$.
The theory corresponding to the cone over $\mathrm{dP}_3$ with $d=12$, for example, has the valencies $T_{\s_B} = (3,3,6)$ and $T_{\s_W} = (4,4,4)$.}
They tell us how many fields appear in each of the terms in the superpotential.
Twice the cycle lengths of $\s_F$ gives the number of edges around the faces of the brane tiling.

\item \textbf{PT-5.}
For physical reasons we restrict attention to the case where $\s_B$ and $\s_W$ have no one-cycles or two-cycles (\textit{i.e.}, the superpotential does not contain tadpoles or mass terms).
We will also restrict to the case where $\s_F$ has no one-cycles. These conditions imply that  there are at least three fields charged under each gauge group.

\ei

\paragraph{Example.}
Let us explicate an illustrative example.
The suspended pinched point (SPP)~\cite{mp,u,Park:1999ep,Feng:2000mi} is defined by
\be
\mathrm{SPP} = \{ (z_1, z_2, z_3, z_4) \in \mathbb{C}^4 ~|~ z_1\, z_2 = z_3\, z_4^2 \} ~.
\ee
The gauge theory associated to the worldvolume of $N$ D$3$-branes at the singularity is captured by the brane tiling depicted in Figure~\ref{fig:spp}.
\bfig[t!]
\bc
\resizebox{0.5\hsize}{!}{
\includegraphics[scale=.4]{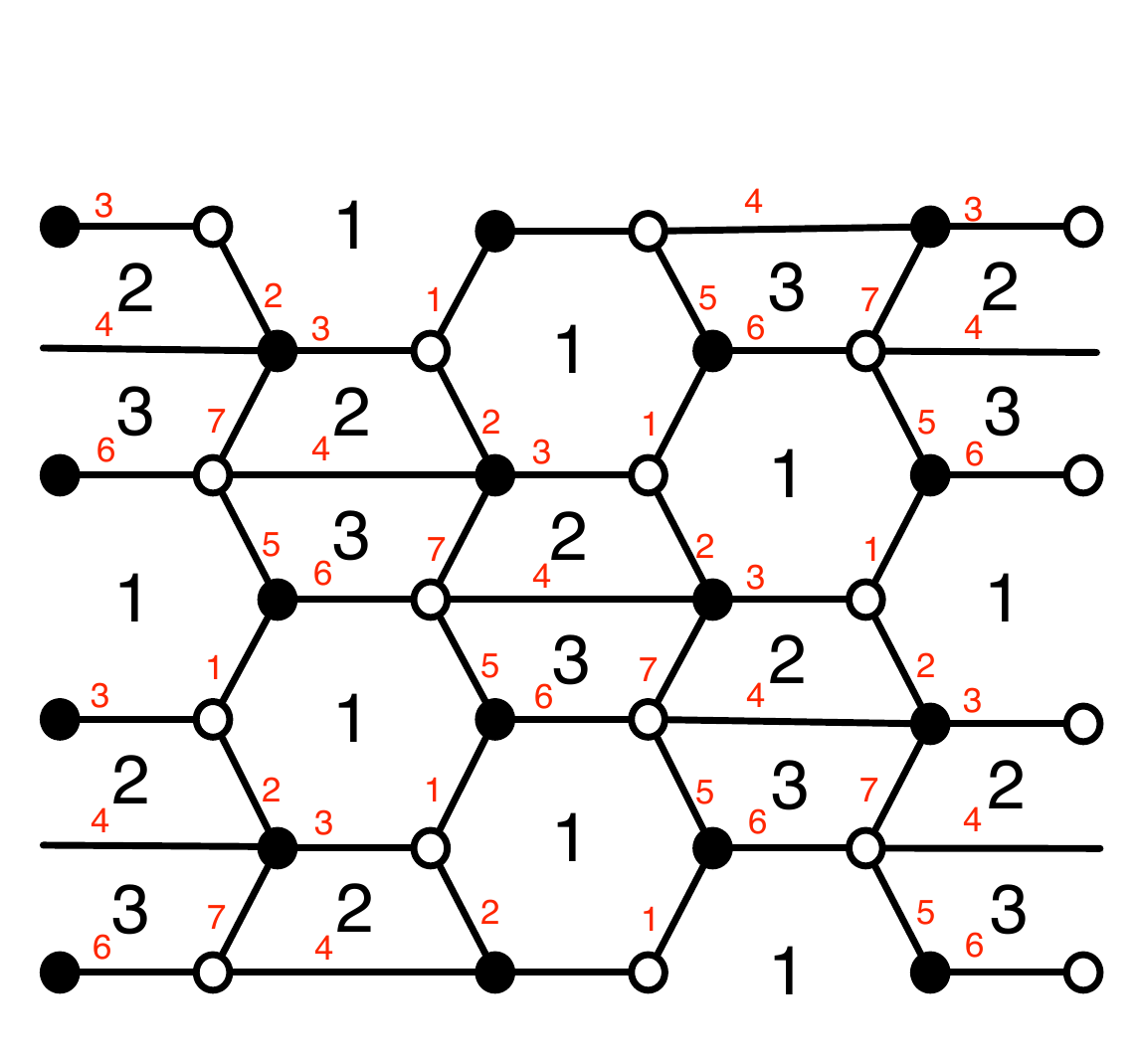}
}
\ec
\caption{
Brane tiling for the suspended pinch point.
The edges are labeled in red, and the gauge groups are labeled in black.
The field content and superpotential interactions are captured by $\sigma_B = (1\ 6\ 5)\ (2\ 4\ 7\ 3)$ and $\sigma_W = (1\ 3\ 2)\ (4\ 7\ 6\ 5)$.
As this is a periodic tiling, faces with the same number are identified.
The dual graph is the extended quiver.
}
\label{fig:spp}
\efig
The superpotential for this brane tiling is given by
\bea
W &=& \mathrm{tr} \left( X_{11} X_{12} X_{21} - X_{11} X_{13} X_{31} + X_{23} X_{31} X_{13} X_{32} - X_{12} X_{23} X_{32} X_{21} \right) \nn \\
&=:& \mathrm{tr} \left( \phi_1 \phi_2 \phi_3 - \phi_1 \phi_6 \phi_5 + \phi_4 \phi_5 \phi_6 \phi_7 - \phi_2 \phi_4 \phi_7 \phi_3 \right) ~.
\eea
The field $X_{ij}$ is a fundamental under gauge group $i$ and an anti-fundamental under gauge group $j$.
We label the edges (fields) in the bipartite graph in red and the faces (gauge groups) in black.
As there are seven fields, we write permutations in the group $S_7$.
The permutation triple that encodes this brane tiling is
\be
\sigma_B = (1\ 6\ 5)\ (2\ 4\ 7\ 3) ~, \qquad
\sigma_W = (1\ 3\ 2)\ (4\ 7\ 6\ 5) ~, \qquad
\sigma_F = (1\ 3\ 5)\ (2\ 7)\ (4\ 6) ~.
\ee
As $C_{\s_B} = C_{\s_W} = 2$ and $C_{\s_F} = 3$,~\eref{eq:rh} is satisfied.
We observe from the figure that correlating with the cycle structure of $\s_F$, there are three faces, two with four edges and one with six edges.
Equivalent descriptions of the brane tiling are obtained through simultaneous conjugation of $(\s_B,\s_W)$.

\subsection{Zig-zag Paths} \label{subsec:zzp}

Information about the toric Calabi--Yau threefold and its toric diagram, as well as the $5$-brane configuration represented by a brane tiling, is contained 
in  cycles on the two-torus with non-trivial winding numbers which are known as \textit{zig-zag paths}~\cite{hv}.
The T-dual of the D$3$-branes probing the toric Calabi--Yau threefold is a brane configuration of D$5$-branes suspended between NS$5$-branes that wrap a holomorphic curve $\Sigma$ in $x,y \in (\mathbb{C}^*)^2$ coordinates as given in Table~\ref{tbranes}.
This holomorphic curve is given by the Newton polynomial in $x,y$ of the toric diagram of the toric Calabi--Yau threefold.
The subspace $(\arg{(x)},\arg{(y)})$ is in $T^2$, where $T^2$ is the two-torus of the brane tiling.
The holomorphic curve in $T^2$ is limited by asymptotic boundaries which correspond to non-trivial cycles on the two-torus known as zig-zag paths.

From the point of view of the bipartite graph on the two-torus given by the brane tiling, the zig-zag paths are constructed as follows~\cite{ks,hv,gsu,fhkpuv}.
Starting from an edge, the zig-zag path turns  right around the adjacent white vertex, crosses the next edge, and turns  left around the following black vertex.
Tracing these steps repeatedly, the resulting path is closed on the torus.

As in~\cite{jrr}, we express the zig-zag path in terms of elements of $S_{2d}$.
We begin by adding labels $+$ and $-$ to each of the edges.
Define a permutation of the set   $ \{ 1^+ , \cdots , d^+ , 1^- , \cdots , d^- \} $ as follows 
\be
{\cal Z}(k^-) = (\sigma_B(k)) ^+ ~, \qquad
{\cal Z}(k^+) = (\sigma_W^{-1}(k))^- ~. \label{eq:zz}
\ee
In shorthand, we write this as ${\cal Z}(\sigma_B, \sigma_W^{-1})$.
We may equivalently write
\be
\Sigma_1 = (\sigma_B)_+  \circ  (\sigma_W^{-1})_-   ~, \qquad \Sigma_2 = (1^+\ 1^-)\ (2^+\ 2^-)\ \ldots\ (d^+\ d^-) ~,
\ee
The $ (\sigma_B)_+ $   permutes the first $d$ elements in  $ \{ 1^+ , \cdots , d^+ , 1^- , \cdots , d^- \} $
while the $ (\sigma_W)_-$ permutes the last $d$ elements  
\bea 
( \sigma_B )_+ (k^+ ) = ( \sigma_B (k) )^+ \qquad \qquad ( \sigma_W )_- ( k^- ) = ( \sigma_W ( k ) )^-   
\eea
The permutation  ${\cal Z}$ can now be written as a product $ { \cal  Z } = \Sigma_2\cdot \Sigma_1 $ : 
\bea
{\cal Z}(k^-) = ( \sigma_B)_+  (   \Sigma_2 (k^-)) = (\sigma_B)_+  ( k^+ ) =   ( \sigma_B (k ) )^+ ~, \cr 
{\cal Z}(k^+) = ( \sigma_W^{-1})_- (\Sigma_2(k^+)) =( \sigma_W^{-1})_- (k^-)  =  ( \sigma_W^{-1}(k) )^- ~.
\eea

\paragraph{Example.}
For SPP, employing~\eref{eq:zz}, we find that
\be
{\cal Z}(\sigma_B, \sigma_W^{-1}) = (1^-\ 3^+\ 7^-\ 6^+)\ (2^-\ 1^+\ 5^-\ 4^+)\ (3^-\ 2^+)\ (4^-\ 7^+)\ (6^-\ 5^+) ~.
\ee
Each cycle in the previous expression is a zig-zag path.
Thus, for the SPP theory, we write
\bea
&& z_1 = (1^-\ 3^+\ 7^-\ 6^+) ~, \quad
z_2 = (2^-\ 1^+\ 5^-\ 4^+) ~, \nn \\
&& z_3 = (3^-\ 2^+) ~, \quad
z_4 = (4^-\ 7^+) ~, \quad
z_5 = (6^-\ 5^+) ~. \label{eq:zzspp}
\eea
Note as well that if we replace $\sigma_W^{-1}$ with $\sigma_W$ in~\eref{eq:zz}, we obtain
\be
{\cal Z}(\sigma_B, \sigma_W) = (1^-\ 5^+\ 6^-\ 1^+\ 2^-\ 3^+)\ (3^-\ 7^+\ 4^-\ 2^+)\ (5^-\ 6^+\ 7^-\ 4^+) ~.
\ee
This yields another permutation element in $S_{2d}$ that tells us which edges circumscribe the faces of the SPP brane tiling.
The numbers that appear in each cycle denote the fields that transform under the gauge groups, and the $\pm$ denotes fundamental and anti-fundamental $U(N)$ indices. Note that the cycle lengths in ${\cal Z}(\sigma_B, \sigma_W)$ are twice the cycle lengths in $\sigma_F=(\sigma_B \sigma_W)^{-1}$.

%----------------------------%
\subsection{Consistent Brane Tilings} \label{subsec:cd}

We are focused on studying brane tilings which correspond to quantum field theories that are \textit{well behaved}. Such quantum field theories flow under the renormalization group to an infrared fixed point that has no accidental global $U(1)$ symmetries. The quiver gauge theory is assumed to be superconformal at this point such that each term in the superpotential~\eref{eq:w} carries $R$-charge $2$. Moreover, the NSVZ beta function takes the form
\be
\beta_i = \frac{3N}{2\left(1-\frac{g_i^2 N}{8\pi^2}\right)} \left[2-\sum_{X \in \partial G_i}(1-R(X))\right] ~,
\ee
where $X\in \partial G_i$ is a bifundamental field charged under the gauge group $G_i$.
The notation is chosen to emphasize that for a brane tiling, $X$ is a brane tiling edge bounding the face corresponding to $G_i$.
The index $i$ therefore runs from $1,\ldots,C_{\s_F}$.
Here, $R(X)$, the $R$-charge of the field $X$, can be obtained by procedures known as $a$-maximization~\cite{Intriligator:2003jj} and volume minimization~\cite{Butti:2005vn,msy1,msy2}.
Accordingly, in order for the quantum field theory corresponding to a brane tiling to be scale invariant, the following conditions need to be satisfied~\cite{fhkvw}
\bea
\sum_{X \in V_a} R(X) = 2 \ \forall\ a ~, \label{eq:rc} \\
\sum_{X \in \partial G_i} (1-R(X)) = 2 \ \forall\ i ~, \label{eq:rcg}
\eea
where $X\in V_a$ are fields that appear in each single trace operator $V_a \subset W$.
There are $C_{\s_B} + C_{\s_W}$ conditions corresponding to~\eref{eq:rc} and $C_{\s_F}$ conditions corresponding to~\eref{eq:rcg}.

We examine a theory for inconsistencies subsequent to performing $a$-maximization.
For some tilings, $a$-maximization yields fields with negative $R$-charges or gauge invariant combinations of fields that violate the unitarity bound.\footnote{
The $R$-charge of a gauge invariant operator must be at least $\frac23$~\cite{ff,dp}.}
The infrared fixed points for such tilings are at present undetermined.
Through examples, we discover that geometrically consistent tilings do not suffer from these pitfalls.

\paragraph{Geometric consistency.}
Any brane tiling which can be tiled on a torus will have locally flat nodes and faces.
This means that the constraints~\eref{eq:rc} and~\eref{eq:rcg} are satisfied.
Of all such tilings, there is a subset for which $a$-maximization generates $R$-charges that respect unitarity.
These are the well behaved quantum field theories that we investigate.
Brane tilings corresponding to such theories obey so called \textit{geometric consistency conditions}.
Similar consistency conditions are discussed in~\cite{b1} and can be connected to various properties of brane tilings linked to zig-zag paths~\cite{2003math.....10326K}, $R$-charges~\cite{k}, Calabi--Yau algebras~\cite{b2}, and cancelation properties of quivers~\cite{2008arXiv0812.4185D,2013arXiv1301.5410B}.

We define geometric consistency in terms of zig-zag paths $z_\alpha$.
A brane tiling is geometrically consistent if in the universal cover the zig-zag paths satisfy the following conditions.
\bi
\item \textbf{CONS-1.}
There are no self-intersecting zig-zag paths. 

\item \textbf{CONS-2.}
If the winding numbers of two zig-zag paths are linearly independent, the zig-zag paths intersect precisely once in the universal cover.

\ei
Functionally, the first condition is easy to verify.
We must simply ensure that a number is not repeated within the $z_\alpha$ obtained from~\eref{eq:zz}~\cite{jrr}.
That is to say, both $k^-$ and $k^+$ cannot appear in any of the cycles in ${\cal Z}(\sigma_B,\sigma_W^{-1})\in S_{2d}$.
Both geometric consistency conditions have been discussed in~\cite{2003math.....10326K,hv,b1,b2}.
In~\cite{b2}, a further condition is added.
This condition states that linearly dependent zig-zag paths cannot intersect at all. 
Together with \textbf{CONS-2} this would imply that all canceling intersections (irrespective of linear dependence) are excluded.
We do not impose the extra condition, and as a result all Seiberg dual phases are consistent by our definitions (\textbf{CONS-1} and \textbf{CONS-2}). 
In the next section we recast geometric consistency in the language of derangements.

%================================%
\section{Consistency and Derangements} \label{sec:cd}

A \textit{derangement} is a permutation without fixed points (for key properties see, for example,~\cite{Cameron}). 
In other words, there are no cycles of length one.
The number of derangements in $S_n$ is the integer nearest to $n!/e$.
These are the \textit{subfactorial numbers}:
\be
!n := n! \sum_{k=0}^n \frac{(-1)^k}{k!} = 1, 0, 1, 2, 9, 44, 265, 1854, 14833, 133496, \ldots ~,
\ee
with generating function
\be
\sum_{n=0}^\infty !n\ x^n = \frac1x \frac{\mathrm{Ei}(1+\frac1x)}{e^{1+\frac1x}} ~,
\ee
where the numerator is written in terms of the exponential integral $\mathrm{Ei}(m) = - \int^\infty_{-m} dt\ \frac{e^{-t}}{t}$.

Let $H$ be the group generated by $\s_B \s_W^{-1}$.
This is an Abelian group.
From above, the first consistency condition is that zig-zag paths have no self-intersections.
The second consistency condition is that two linearly independent zig-zag paths do not intersect more than once while being oriented from one intersection point to the next in the same way.
Let us restate these conditions in terms of group theory.
\bi
\item \textbf{CONS-1$'$.}
For any $h \in H $, the permutation $h \s_B $ has no one-cycle.
That is to say, $h \s_B$ is a \textit{derangement}.

\item \textbf{CONS-2$'$.}
Choose all possible pairs $h_1, h_2 \in H$.
If the permutation $h_1 \s_B h_2 \s_W$ is not a derangement, the fields in the one-cycles belong pairwise to zig-zag paths that are linearly dependent.

\ei
Failure to satisfy these conditions indicates a lack of geometric consistency. The simpler condition that $h_1 \s_B h_2 \s_W$ is a derangement amounts to excluding all canceling intersections between pairs of zig-zag paths. 
In this section, we explicate how these conditions are equivalent to the ones quoted previously.

\bfig[h!]
\bc
\resizebox{0.9\hsize}{!}{
\includegraphics[scale=.6]{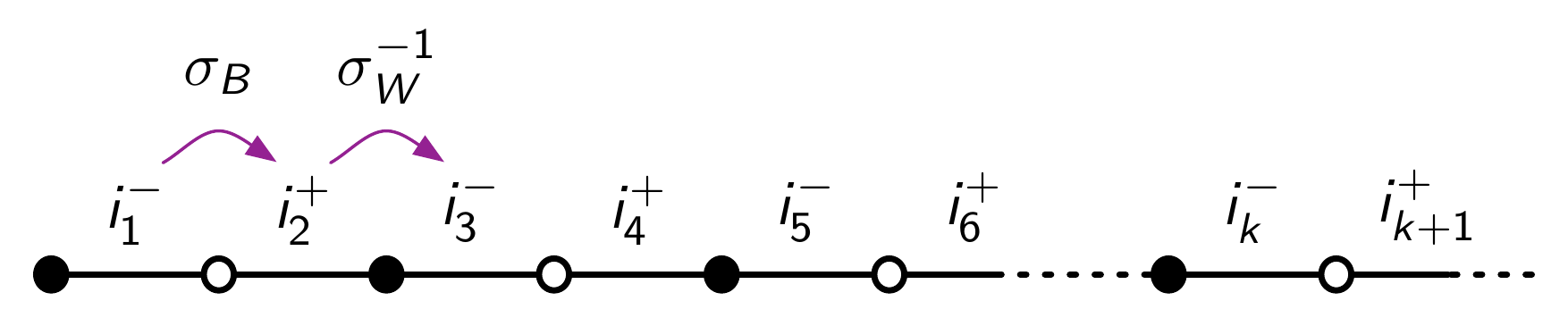}
}
\ec
\caption{
 A zig-zag path.
\label{fig:cons-1}}
\efig
Figure~\ref{fig:cons-1} illustrates a zig-zag path.
The absence of self-intersections means that no label appears twice within a zig-zag path.
In particular, $i_j \ne i_k$ for $j\ne k$.
If two labels were to be the same --- that is, when a zig-zag path self-intersects --- the labels appear with opposite signs.
Without loss of generality, assume that $i_1^- = i_{k+1}^+$ for some $k$.
To go from $i_1^-$ to $i_{k+1}^+$ we must apply $(\sigma_B \sigma_W^{-1})^\frac{k-1}{2} \sigma_B$.
This means that
\be
(i_1) h \sigma_B = (i_{k+1}) = (i_1) ~, \label{eq:c1}
\ee
for some $h\in H$.\footnote{We have written~\eref{eq:c1} in terms of left actions.}
Since there is a fixed point, $h\sigma_B$ is a permutation with a one-cycle.
Thus, requiring that for any $h\in H$, $h\sigma_B$ is a derangement negates the possibility of self-intersection.

\bfig[h]
\bc
\resizebox{0.9\hsize}{!}{
\includegraphics[scale=.5]{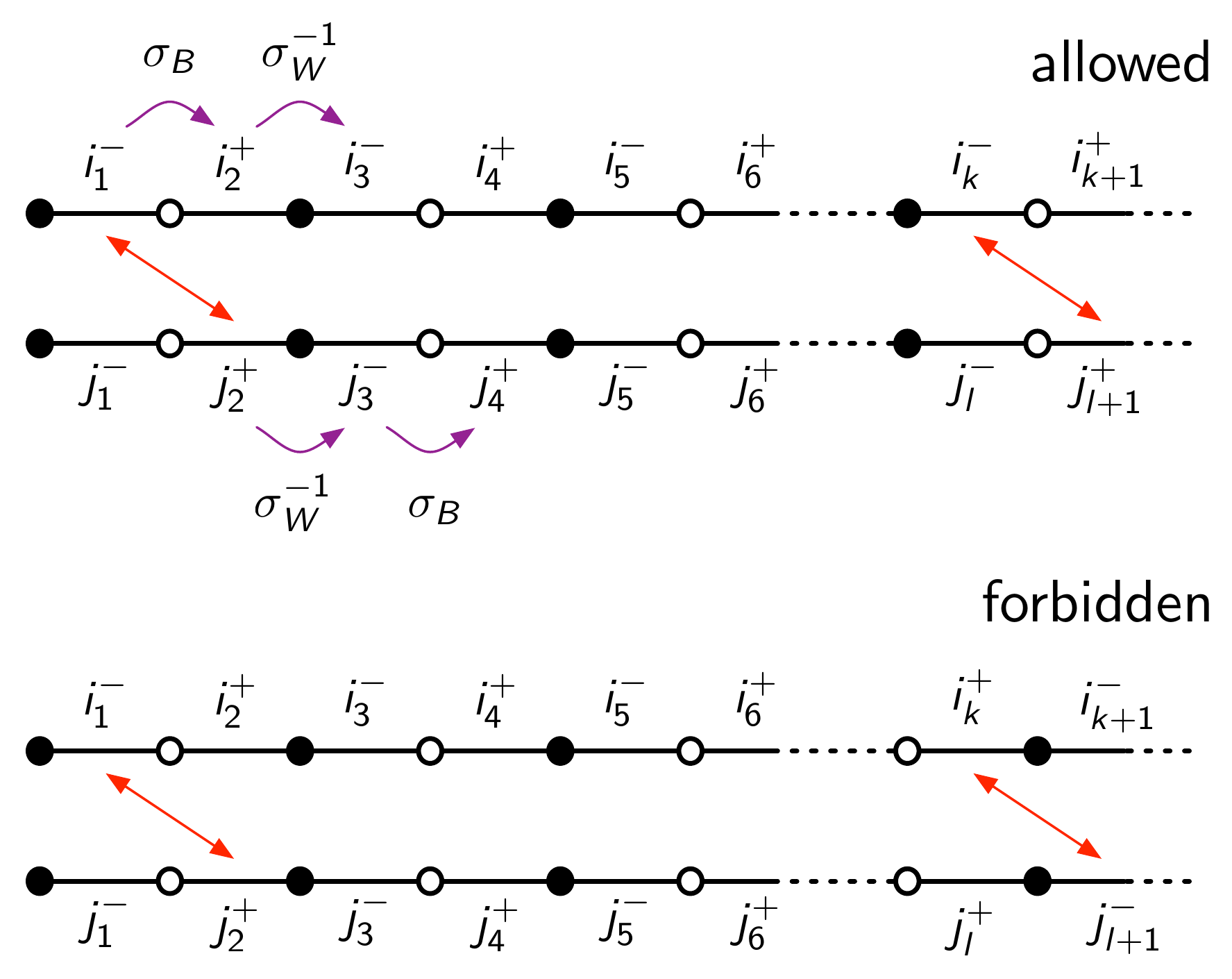}
}
\ec
\caption{
A pair of independent zig-zag paths.
Labels connected by red arrows are the same.
If $i_1^-$ and $j_2^+$ represent the same field, $i_k^+$ and $j_{\ell+1}^-$ cannot be the same field for any $k,\ell$.
However, as in the case of $\mathbb{F}_0 (\mathrm{II})$, $i_k^-$ and $j_{\ell+1}^+$ can be the same field.
}
\label{fig:cons-2}
\efig
Figure~\ref{fig:cons-2} illustrates a pair of linearly independent zig-zag paths.
Without loss of generality, we have $i_1^-$ and $j_2^+$ representing the same field.
In this case, any second intersection of the two zig-zag paths requires that some minus label from the upper zig-zag path is the same as some plus label from the lower zig-zag path.
Parity is preserved in the intersections.
The forbidden configurations have the parity reversed:
there is an $i_k^+$ that represents the same field as $j_{\ell+1}^-$.
Thus,
\bea
(i_k) = (i_1) (\sigma_B \sigma_W^{-1})^\frac{k-1}{2} \sigma_B &=& (j_{\ell+1}) = (j_2) \sigma_W^{-1} (\sigma_B \sigma_W^{-1})^\frac{\ell-2}{2} \nn \\
\Longrightarrow \quad (i_1) h_1 \sigma_B h_2 \sigma_W &=& (i_1) ~. \label{eq:repeat}
\eea
Thus, $h_1 \sigma_B h_2 \sigma_W$ has a one-cycle for some $h_1, h_2\in H$.
In order to avoid such an intersection, we require that $h_1 \sigma_B h_2 \sigma_W$ be a derangement for all $h_1, h_2\in H$.

The situation in~\eref{eq:repeat} can occur if it happens that the zig-zag paths under consideration are not independent.
That is to say, if the two zig-zag paths are proportional to each other, then for some elements $h_1, h_2\in H$, the product $h_1 \sigma_B h_2 \sigma_W$ may not be a derangement.
The one-cycles then identify the fields that are common to both of the zig-zag paths.

\paragraph{Example 1.}
The zig-zag paths for SPP are quoted in~\eref{eq:zzspp}.
We can construct the intersection matrix
\be
I_\mathrm{SPP} = \langle z_a, z_b \rangle = \left( \ba{ccccc} 0 & 1 & -1 & 1 & -1 \cr -1 & 0 & 1 & -1 & 1 \cr 1 & -1 & 0 & 0 & 0 \cr -1 & 1 & 0 & 0 & 0 \cr 1 & -1 & 0 & 0 & 0 \ea \right) ~.
\ee
Following the conventions of~\cite{jrr}, when $i^-$ appears in $z_a$ and $i^+$ appears in $z_b$, this contributes $+1$ to the intersection matrix element $I_{ab}$.
Similarly, when $i^+$ appears in $z_a$ and $i^-$ appears in $z_b$, this contributes $-1$ to the intersection matrix element $I_{ab}$.
It is clear that $I_{ab} = -I_{ba}$.
Since we can express the zig-zag paths in the homology of $\mathbb{T}^2$ as linear combinations of the $a$-cycle and $b$-cycle of the torus, the intersection matrix has rank two.
By inspection, $z_3$, $z_4$, and $z_5$ are not linearly independent.
As they have no fields in common, these zig-zag paths do not intersect.
Noting that $h := \sigma_B \sigma_W^{-1} = (1\ 7)\ (2\ 5)\ (3)\ (4)\ (6)$, we have an Abelian group $H = \{1, h\}$.
That the conditions~\textbf{CONS-1$'$} and~\textbf{CONS-2$'$} are satisfied can be verified by explicit computation.

\bfig[h]
\bc
\resizebox{0.6\hsize}{!}{
\includegraphics[scale=.5]{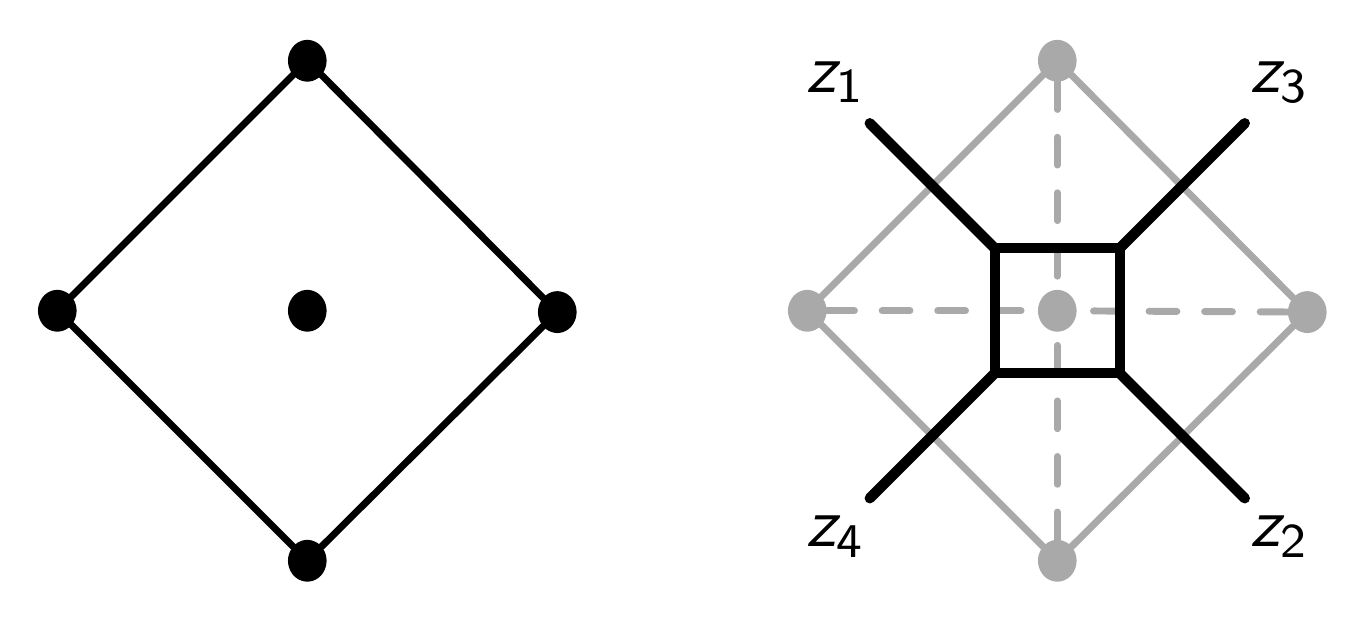}
}
\ec
\caption{
The toric diagram of $\mathbb{F}_0$ and the corresponding dual $(p,q)$-web diagram.
The external legs of the $(p,q)$-web diagram, which we label $z_1,z_2,z_3,z_4$, correspond to zig-zag paths whose winding numbers on the two-torus are given by $(p,q)$.
}
\label{ftoricf0}
\efig

\paragraph{Example 2.}
The cone over the zeroth Hirzebruch surface $\mathbb{P}^1\times \mathbb{P}^1$ is known as $\mathbb{F}_0$.
There are two Seiberg dual phases of $\mathbb{F}_0$~\cite{Feng:2001xr,fhkvw}.
The first phase corresponds to a brane tiling which contains four gauge groups and eight fields.
The second phase is a brane tiling which contains four gauge groups and twelve fields.
Both phases have zig-zag paths that relate to the dual $(p,q)$-web diagram of the toric diagram of $\mathbb{F}_0$ as shown in Figure~\ref{ftoricf0}.
The winding numbers of the zig-zag paths directly relate to the directions of the external legs of the $(p,q)$-web diagram.
In both phases, the set of zig-zag paths with the same winding numbers intersect in different ways in order to produce the different quivers associated to the two phases.
As it is the case for dual brane tilings, there is always a minimal case in which zig-zag paths intersect at most once.
The minimal case is phase I of $\mathbb{F}_0$.
We will focus on the second phase, $\mathbb{F}_0(\mathrm{II})$ as depicted in Figure~\ref{fig:f0ii}.
\bfig[h]
\bc
\resizebox{0.6\hsize}{!}{
\includegraphics[scale=.5]{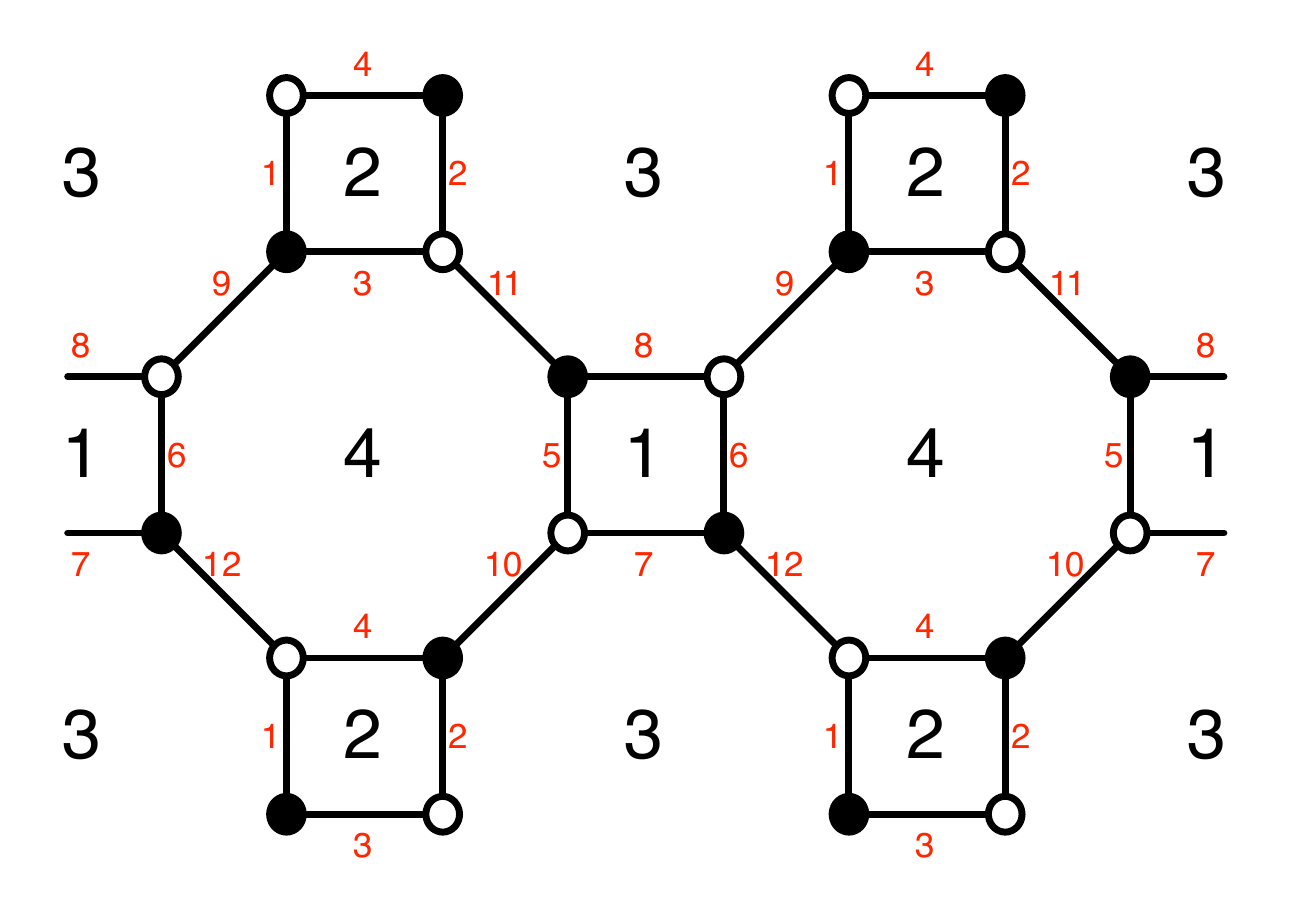}
}
\ec
\caption{
The brane tiling for $\mathbb{F}_0(\mathrm{II})$.
}
\label{fig:f0ii}
\efig

We note that for $\mathbb{F}_0(\mathrm{II})$, the permutation tuples encoding the superpotential are
\be
\sigma_B = (1\ 9\ 3)\ (2\ 10\ 4)\ (5\ 8\ 11)\ (6\ 7\ 12) ~, \qquad
\sigma_W = (1\ 4\ 12)\ (2\ 3\ 11)\ (5\ 10\ 7)\ (6\ 9\ 8) ~.
\ee
From this, using~\eref{eq:zz}, we compute the zig-zag paths
\bea
z_1 = (1^-\ 12^+\ 6^-\ 8^+\ 11^-\ 3^+) ~, && z_2 = (2^-\ 11^+\ 5^-\ 7^+\ 12^-\ 4^+) ~, \nn \\
z_3 = (3^-\ 2^+\ 10^-\ 5^+\ 8^-\ 9^+) ~, && z_4 = (4^-\ 1^+\ 9^-\ 6^+\ 7^-\ 10^+) ~.
\eea
We observe that $z_1\cap z_3$ at fields $3$ and $8$.
Both of these fields carry plus labels in $z_1$ and minus labels in $z_3$.
The intersections $z_1\cap z_4$, $z_2\cap z_3$, and $z_2\cap z_4$ have a similar structure.
This is what we expect from~\textbf{CONS-2$'$}.

Notice, however, $z_1\cap z_2$ and $z_3\cap z_4$.
In the former $z_1\supset 11^-, 12^+$ and $z_2\supset 11^+, 12^-$, and in the latter $z_3\supset 9^+, 10^-$ and $z_4\supset 9^-, 10^+$.
Noting the signs, the zig-zag paths intersect twice with opposite parity.
At first glance, this appears to contradict the geometric consistency condition.
There is no inconsistency, however, because the pairs $(z_1, z_2)$ and $(z_3, z_4)$ are not linearly independent.
The intersection matrix for $\mathbb{F}_0(\mathrm{II})$ is
\be
I_{\mathbb{F}_0(\mathrm{II})} = \langle z_a, z_b \rangle = \left( \ba{cccc} 0 & 0 & -2 & 2 \cr 0 & 0 & 2 & -2 \cr 2 & -2 & 0 & 0 \cr -2 & 2 & 0 & 0 \ea \right) ~.
\ee
We compute the null space of $I_{\mathbb{F}_0(\mathrm{II})}$ as $(1,1,0,0)^T$ and $(0,0,1,1)^T$, which explains that $z_1 + z_2 = z_3 + z_4 = 0$.
Unlike the previous example of SPP, all of the zig-zag paths intersect each other.
When the zig-zag paths are linearly dependent, we can have intersections of opposite orientation without violating geometric consistency.
We also note that when zig-zag paths are linearly dependent, they correspond to $(p,q)$-legs in the $(p,q)$-web diagram that are parallel to each other as shown in Figure~\ref{ftoricf0}.

The group generated by
\be
h:= \sigma_B \sigma_W^{-1} = (1\ 6\ 10)\ (2\ 5\ 9)\ (3\ 12\ 8)\ (4\ 11\ 7)
\ee
consists of three elements: $H = \{1, h, h^2\}$.
We may check that~\textbf{CONS-1$'$} holds, which corroborates the fact that the zig-zag paths do not self-intersect.
To verify~\textbf{CONS-2$'$}, we must examine nine possible combinations $h_1 \sigma_B h_2 \sigma_W$ with $h_1, h_2\in H$.
Eight of these are derangements.
The lone exception is $h_1 = h_2^{-1} = h$, for which the product
\be
h_1 \sigma_B h_2 \sigma_W = (1\ 2)\ (3\ 4)\ (5\ 6)\ (7\ 8)\ (9)\ (10)\ (11)\ (12) \label{eq:notderanged}
\ee
contains one-cycles.
This is a flag that zig-zag paths proportional to each other intersect with opposed parities.
In fact, the four one-cycles in~\eref{eq:notderanged} identify precisely those fields in the intersections $z_1\cap z_2$ and $z_3\cap z_4$.
We know that these zig-zag paths are proportional to each other from examination of the intersection matrix.

\paragraph{Example 3.}
After $\mathbb{F}_0$, del Pezzo $2$ is the next model which exhibits two dual brane tilings~\cite{Feng:2001xr,Feng:2002zw,Hanany:2012hi}.
The first phase of del Pezzo $2$ has eleven fields and is the model where the zig-zag paths of the brane tiling intersect minimally.
The second phase of del Pezzo $2$, on which we concentrate in this example, is a theory with thirteen fields.
For this phase, $h_1 \sigma_B h_2 \sigma_W$ fails to be a derangement because zig-zag paths that intersect twice are proportional to each other.
We can verify this starting from the tuples
\be
\sigma_B = (1\ 4\ 13)\ (2\ 5\ 8\ 11)\ (3\ 10\ 9)\ (6\ 12\ 7) ~, \quad
\sigma_W = (1\ 3\ 2)\ (4\ 6\ 5)\ (7\ 9\ 8)\ (10\ 13\ 12\ 11) ~.
\ee
The cases in which~\textbf{CONS-2$'$} is subtle correspond to the Seiberg dual phases of ${\cal N}=1$ theories with more than the minimum number of fields.
In general, in the case of a non-minimal phase, linearly dependent zig-zag paths that intersect multiple times correspond to parallel $(p,q)$-legs in the dual of the toric diagram.

In this manner, following a straightforward recipe, we demonstrate the geometric consistency of various brane tilings using nothing more than the permutation tuples that define superpotential interactions.
Coding this, we are able to verify the methodology presented for all conceivable superpotentials with fields $d\le 10$ that satisfy the permutations and tuples conditions quoted above.

%================================%
\section{Counting and Future Directions } \label{sec:p}

A pair of permutation tuples provides a compact terminology for expressing the matter content and the interactions of a large class of ${\cal N}=1$ theories in four dimensions given by brane tilings.
As noted in~\cite{jrr,hhjprr1}, brane tilings are dessins d'enfants, which supply combinatorial invariants for the action of the absolute Galois group of the rational numbers.
In this note, we have established that geometric consistency conditions for brane tilings can be written purely in terms of the properties of the tuples that describe the terms in the superpotential of the field theory.
We simply consider an Abelian group $H ( \sigma_B  \sigma_W^{-1}  ) $ generated by $\sigma_B\sigma_W^{-1}$ and use the elements $h$ in this group to generate certain permutations in $S_d$ of the form $h \sigma_B $ and $h_1\sigma_B h_2\sigma_W$.
Permutations of the first type must always be derangements.
Permutations of the second type can fail to be derangements only when zig-zag paths are proportional to each other.
Checking these conditions ensures the consistency of a brane tiling.

Fixing the number of chiral fields $d$, it is possible to encode a simple algorithm to consider pairs of elements of $S_d$ consonant with the permutations and tuples conditions and verify that $(\sigma_B, \sigma_W)$ satisfy the geometric consistency conditions as well.
This provides a method for generating brane tilings with an arbitrary number of edges.

In fact, following our analysis of the geometric consistency conditions in terms of permutation tuples, we can write down counting formulas for the 
number of brane tilings for a given number of chiral fields $d$, which satisfy the conditions related to  derangements. These formulae also exclude 
tilings where linearly dependent zig-zags have canceling intersection, so they give a lower bound on the geometrically consistent tilings. This lower bound counts the number of universality classes of infrared fixed points for the brane tilings. That is to say, the formulas count toric diagrams.
Let $T_W$,  $T_B$, and $T_F$  be three conjugacy classes of $S_d$ such that $\sigma_W \in T_W$, $\sigma_B \in T_B$, and 
$ \sigma_F = ( \sigma_B \sigma_W)^{ -1} \in T_F$.
Note that the choice of $T_W$, $T_B$, $T_F$ should be such that they satisfy the properties \textbf{PT-1} to \textbf{PT-5} outlined in Section~\ref{subec:bt}.
For instance, in order for the brane tilings to be on $T^2$, the conjugacy classes should be such that $d - C_{\s_B} - C_{\s_W} - C_{\s_F} = 0$.

The number of tilings satisfying the derangement conditions,  counted with inverse automorphisms,  is 
\bea\label{ecountA}
N_{inv}  (d;T_B ,T_W,T_F) &=&
\frac{1}{d!} 
\sum_{\sigma_B  \in T_B }
\sum_{\sigma_W \in T_W }
\sum_{\sigma_F \in T_F }
\delta(\sigma_B  \sigma_W \sigma_F )
~~
\nn\\
&&
\prod_{h_1, h_2 , h_3  \in H(\sigma_B\sigma_W^{-1} )}
\sum_{\alpha_1 \in D_d}
\sum_{\alpha_2 \in D_d}
\delta(h_1 \sigma_B  \alpha_1 )
\delta(h_2  \sigma_B h_3 \sigma_W  \alpha_2)
~,~
\nn\\
\eea
where $D_d$ is the set of derangements.
This is a symmetry weighted counting that includes also disconnected bipartite graphs.
In order to include connected bipartite graphs as brane tilings one needs to select only $\sigma_B ,\sigma_W $ which generate a transitive group.
For further explanation of the symmetry factors and applications of delta functions over symmetry groups 
in  the context of counting of bipartite graphs, see~\cite{deMelloKoch:2011uq,deMelloKoch:2012tb}. 
An alternative counting is over equivalence classes of  permutation tuples,   with weight one, the so called unweighted counting, which gives the sum
\bea\label{ecountB}
N(d;T_B ,T_W,T_F ) &=& \frac{1}{d!} 
\sum_{\gamma\in S_d} 
\sum_{\sigma_B  \in T_B }
\sum_{\sigma_W \in T_W}
\sum_{\sigma_F  \in T_W}
\delta(\sigma_B \sigma_W \sigma_F )\delta(\gamma\sigma_B  \gamma^{-1} \sigma_B^{-1})
\delta(\gamma\sigma_W \gamma^{-1} \sigma_W^{-1})
\nn\\
&&
\hspace{2cm} 
\prod_{h_1, h_2 , h_3 \in H(\sigma_B \sigma_W^{-1} )} 
\sum_{\alpha_1 \in D_d}
\sum_{\alpha_2 \in D_d}
\delta(h_1 \sigma_B   \alpha_1)
\delta(h_2 \sigma_B h_3 \sigma_W \alpha_2) 
~.~
\nn \\
\eea
This counts equivalence classes of  permutation pairs, obeying the derangement conditions associated with zig-zag intersections, 
 without inverse automorphisms. This is achieved  by summing over an extra permutation $\gamma$ using Burnside's Lemma.
As before, the sum over $\sigma_B,\sigma_W$ needs to be restricted to $\sigma_B,\sigma_W$ which generate a transitive group in order to make sure that the sum counts connected bipartite graphs corresponding to brane tilings. By summing over all $T_W,T_B,T_F$ satisfying the properties \textbf{PT-1} to \textbf{PT-5} outlined in Section \ref{subec:bt}, the counting in~\eref{ecountA} and \eref{ecountB} for some initial values of $d=1,\dots,6$ is summarized as follows:
\begin{table}[ht!!]
\centering
\begin{tabular}{c|ccccccc}
$d$ & 1 & 2 & 3 & 4 & 5 & 6 & $\dots$ \\
\hline
$N_{inv}(d)$ & 0 & 0 & 1/3 & 1/4 & 0 & 1/2 & $\dots$\\
$N(d)$ & 0 & 0 & 1 & 1 & 0 & 1 & $\dots$\\
\end{tabular}
\caption{Number of consistent tilings for $d$ fields.}
\label{tninv}
\end{table}

\noindent
Note that there is precisely one consistent brane tiling with three fields (${\cal N}=4$ super-Yang--Mills theory), one consistent brane tiling with four fields (the conifold), and one consistent brane tiling with six fields ($\mathbb{C}^2/\mathbb{Z}_2\times \mathbb{C}$).
There are no geometrically consistent brane tilings with five fields.
This is what $N(d)$ tells us.
The denominator in $N_{inv}(d)$ reproduces the order of the automorphism group of the defining permutation tuples --- \textit{i.e.}, the number of elements in $S_d$ that, acting through conjugation, map the pair $(\sigma_W, \sigma_B)$ to itself.

Work in progress studies the implications of geometric consistency and its failure on the underlying brane configurations for the brane tilings as well as the dynamics of the corresponding supersymmetric gauge theories.
In particular, we aim to investigate the CFT fixed point in the infrared to which a geometrically inconsistent brane tiling will flow under the renormalization group.
Given the interplay between supersymmetric gauge theories and dessins d'enfants, and the mathematical  interest in  absolute Galois group actions on the latter~\cite{Schneps}, 
this work  also raises the interesting question of whether  bipartite graphs restricted  by  the derangement conditions  specified here 
form closed orbits of the Galois group action.

\section*{Acknowledgements}
We are grateful to Yang-Hui He for collaboration during early stages of this work.
We thank Stefano Cremonesi, Jurgis Pasukonis, and Diego Rodriguez-Gomez for conversations on related topics.
We  acknowledge the International Centre for Mathematical Sciences in Edinburgh for hosting the ``Quivers, Tilings, and Calabi--Yaus'' workshop in 2014, where some of the results reported in this paper were obtained, and we thank the participants at the meeting for discussions.
We also acknowledge the summer workshops in 2014 and 2015 at the Simons Center for Geometry and Physics which gave an opportunity to progress this project.
A.~H.~ is supported by 
STFC Consolidated Grant ST/J0003533/1, and
EPSRC Programme Grant EP/K034456/1.
VJ is supported by the National Research Foundation of South Africa and the South African Research Chairs Initiative. SR is supported by STFC Grant ST/J000469/1, String Theory, Gauge Theory, and Duality.

%========================================%
%========================================%
\bibliographystyle{JHEP}
\bibliography{mybib}

\end{document}